\documentclass{iau}

\usepackage{graphicx}
\usepackage{amsmath}
\usepackage{hyperref}
\usepackage{threeparttable}
\usepackage{cleveref}

\newcommand{\splitcell}[2]{%
\begin{tabular}{@{}c@{}c@{}c@{}}
#1 & \, to \, & #2
\end{tabular}%
}

\begin{document}

\lefttitle{M\'endez \textit{et al.}}
\righttitle{The Ohio SETI Program}

\jnlPage{1}{7}
\jnlDoiYr{2026}
\doival{10.1017/xxxxx}

\aopheadtitle{Proceedings IAU Symposium}
\editors{J. Haqq-Misra \&  R. Kopparapu, eds.}

\title{The Ohio SETI Program – The Last Decades}

\author{Abel M\'endez$^1$, Robert S. Dixon$^2$, and Russell K. Childers$^3$}

\affiliation{$^1$Planetary Habitability Laboratory (PHL), University of Puerto Rico at Arecibo}
\affiliation{$^2$Former Acting Director, Ohio State University Radio Observatory}
\affiliation{$^3$Former Chief Observer, Ohio State University Radio Observatory}

\begin{abstract}

The Ohio State University Radio Observatory (OSURO), known as the Big Ear, played a pivotal role in both radio astronomy and the Search for Extraterrestrial Intelligence (SETI). Following the completion of the Ohio Sky Survey, the facility was repurposed in 1973 as the world’s first full-time dedicated SETI observatory and operated continuously until its decommissioning in 1998. During this period, the Ohio SETI Program evolved from an 8-channel hydrogen-line receiver into increasingly sophisticated survey systems. Over three decades, these surveys covered approximately 70\% of the radio sky using a largely consistent instrumental configuration, creating one of the most extensive long-term radio astronomy archives ever assembled. The program is best known for the detection of the Wow! Signal in 1977, but it also accumulated an archive of over 40,000 transient narrowband events, revealed unusual concentrations of radio bursts near the Galactic Center, and established one of the longest continuous radio monitoring records in astronomy. Following the closure of the Big Ear, its scientific legacy continued through Project Argus and, more recently, the Arecibo Wow! project. This paper provides an overview of the final decades of the Ohio SETI Program, including its instrumentation, survey strategies, scientific discoveries, and enduring impact on SETI, time-domain radio astronomy, and the preservation of historical astronomical data. Despite its scientific significance, most of the data collected by the Ohio SETI Program remains unexplored, leaving a unique archive available for future research.

\end{abstract}


\maketitle

\section{Introduction}
\begin{figure}
    %\centering
    \includegraphics[width=1\linewidth]{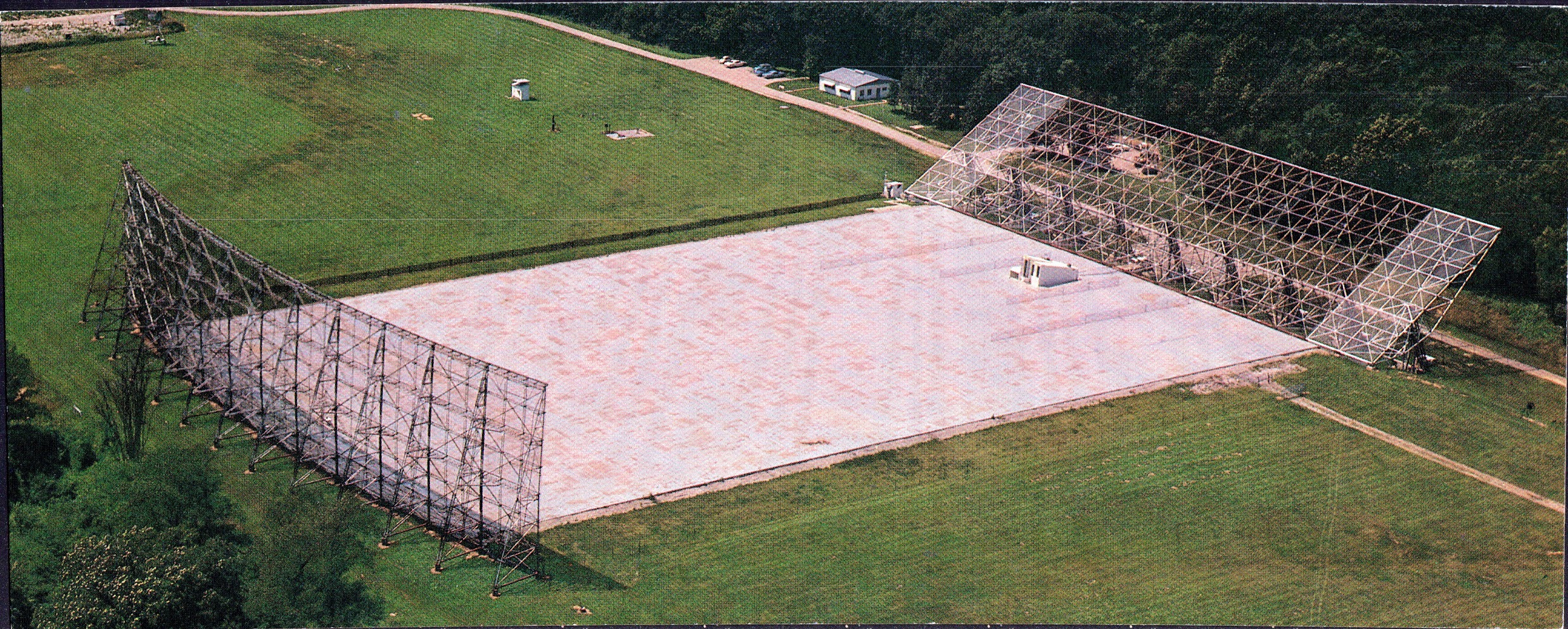}
    \caption{The Ohio State University ``Big Ear" Radio Observatory (OSURO).}
    \label{fig:bigear}
\end{figure}

The Ohio State University Radio Observatory (OSURO), best known as the Big Ear, was a unique and pioneering facility in the history of radio astronomy \citep{1995betl.book.....K,bigear}. The instrument employed a Kraus-style telescope configuration, providing a physical collecting area of 2,200 m$^2$, equivalent to that of a circular aperture with a diameter of 53 m, and was operated as a meridian-transit instrument (\Cref{fig:bigear}). Originally designed for the Ohio Sky Survey (OSS), this instrument generated, for its time, the most extensive all-sky catalog of natural radio sources, comprising nearly 20,000 detections, and enabled the identification of some of the most distant objects then known in the Universe \citep{1977VA.....20..445K}.

As the initial funding ended in 1972, the observatory faced the very real possibility of obsolescence and abandonment, as no immediate funding was available to keep it running. The publication of NASA's Project Cyclops report in 1971 highlighted the deep theoretical need for dedicated, long-term observational searches for extraterrestrial intelligence (SETI) \citep{1973Icar...19..425O,1971asee.nasa.....O}. Recognizing that the Big Ear was uniquely suited for large-scale, continuous sky surveys, Robert S. Dixon, with the support of Director John D. Kraus, transitioned the facility in 1973 into the world's first full-time dedicated SETI program \citep{dixon1973,dixon1977}. This transition marked the beginning of a decades-long pursuit that fundamentally altered the observatory's scientific mission.

Despite its groundbreaking work, including the detection of the famous Wow! Signal in 1977, the existence of the observatory was severely threatened in 1983 \citep{1995betl.book.....K}. The land beneath and surrounding the telescope was unexpectedly sold to a developer with explicit intentions to tear the telescope down and completely remove it to expand a neighboring golf course. This impending destruction ignited a massive public outcry and garnered extensive coverage in the international press. Through a tremendous collaborative struggle involving scientists, the public, and university officials, the telescope was temporarily saved, culminating in a long-term lease that allowed SETI operations to continue.

In the wake of this near-closure, and facing the reality that the university could not provide funding for the observatory staff, the North American AstroPhysical Observatory (NAAPO) was established in 1983 \citep{naapo}. Formed as a non-profit benefactor organization under the direction of Philip Barnhart from Otterbein College, NAAPO became the sustaining force of the observatory. It successfully coordinated an enthusiastic and diverse staff of volunteers, ranging from professional engineers and astronomers to computer programmers and local students. Due to this dedicated volunteer force, the facility was able to maintain its observations and implement technological upgrades.

Despite the valiant efforts of NAAPO volunteers and the continued scientific productivity of the observatory, the encroaching pressures of commercial land development ultimately proved insurmountable. In 1998, as the extended lease on the land expired, the Big Ear was finally decommissioned and dismantled \citep{1998AAS...193.1117B}. Although the physical structure was lost due to the expansion of the adjacent golf course, the scientific legacy, comprehensive data archives, and pioneering methodologies developed during its tenure firmly established the Ohio SETI Program as a foundational pillar in the history of SETI.

\section{Surveys}

The Ohio SETI Program can be divided into five distinct phases, each characterized by specific survey strategies and instrumental configurations (\Cref{tab:surveys}). The survey methodology primarily employed a drift-scan mode, in which the telescope was fixed at a constant declination for extended intervals, typically a minimum of three days, before being incrementally repositioned in discrete steps of 1/3 degrees in declination. Beam-switching techniques were implemented to acquire two adjacent measurements for each sky position, thereby enabling improved discrimination between celestial signals and suppressing terrestrial radio-frequency interference. Interruptions were necessary for the installation or repair of equipment, maintenance, and calibration procedures. Repeated observations at identical declinations improved data redundancy and facilitated independent verification of candidate signal detections. In later stages of the program, instrumental modifications were introduced to allow about one hour of tracking on selected sky positions.

\begin{table*}[ht]
\small
\centering
\begin{threeparttable}
\caption{Summary of the Ohio SETI Program survey phases.}
\begin{tabular}{ccccccc}
\hline
Phase &
Survey Period &
Channels &
Frequency (MHz)&
Bandwidth &
Time Step &
Coverage* \\

\hline

I &
1973 -- 1976 &
8 &
$\sim$1418 -- 1423 &
20, 50, 100 kHz &
10 s &
\splitcell{$+14^\circ$}{$+48^\circ$} \\
%$+14^\circ$ -- $+48^\circ$ \\

II &
1977 -- 1984 &
50 &
$\sim$1418 -- 1423 &
10 kHz &
12--15 s &
\splitcell{$-36^\circ$}{$+63^\circ$} \\
%$-36^\circ$ -- $+63^\circ$* \\

III &
1993 -- 1997 &
3000 &
1400 -- 1700 &
100 kHz &
20 s &
\splitcell{$-36^\circ$}{$+63^\circ$} \\
%$-36^\circ$ -- $+63^\circ$ \\

IV &
1995 -- 1997 &
4M+ &
1421 -- 1424 &
0.6 Hz &
1.7 s &
\splitcell{$-25^\circ$}{$+02^\circ$} \\
%$-25^\circ$ -- $+2^\circ$ \\

V &
2003 -- 2025 &
13k+ &
$\sim$1420 &
4.77 Hz &
209 ms &
All-sky \\

\hline
\end{tabular}

\label{tab:surveys}

\begin{tablenotes}
\footnotesize
\item * Not all declinations were observed, or observed uniformly.
\end{tablenotes}

\end{threeparttable}
\end{table*}

\subsection{Phase I - 1973–1976}

The initial phase of the Ohio SETI Program relied on an 8-channel narrowband receiver system originally built by Bill Brundage for 21-cm galactic hydrogen-line observations \citep{dixon1973, cole1976, dixon1977}. The channels were centered near the 1420.406 MHz rest frequency of hydrogen, Doppler corrected to the Galactic Standard of Rest (GSR). These channels possessed varying bandwidths ranging from 20 to 100 kHz, gradually widening as their frequency distance from the center channel increased. Data processing during this era was entirely analog and manual; the output of the eight channels was plotted as continuous line traces on mechanical pen-chart recorders. These paper charts were laboriously scanned by the eye to isolate transient radio anomalies or signals of interest.

\subsection{Phase II - 1977–1984}

In 1975, the program underwent a major upgrade by incorporating a 50-channel filter bank receiver on loan from the National Radio Astronomy Observatory (NRAO) \citep{1985IAUS..112..305D}. Each channel possessed a narrowband bandwidth of 10 kHz, which yielded a total instantaneous monitoring bandwidth of 500 kHz. The system utilized a two-stage GasFET preamplifier system that achieved an overall system noise temperature of approximately 100 K.

Data processing was largely automated through an IBM 1130 computer running custom machine code and FORTRAN IV. The software continuously commanded a computer-controlled frequency synthesizer (serving as the second local oscillator) to actively Doppler-correct the center frequency back to the GSR. The computer sampled each of the 50 channels once per sidereal second, discarded aberrant noise spikes, and averaged the remaining data over 10-second integration blocks. It dynamically subtracted a moving baseline and normalized the remainder against the standard deviation ($\sigma$) of the background noise.

Simultaneously, a matched-filter routine calculated the cross-correlation between the received power and the known dual-beam antenna pattern. Detections that exceeded $3\sigma$, matched the antenna pattern, and did not span multiple adjacent channels were declared valid celestial sources and permanently preserved on punched cards.

\subsection{Phase III - 1993–1997 (LOBES)}

The Low-Budget Extraterrestrial-Intelligence Search (LOBES) transitioned from the legacy IBM computing platform to a Digital Equipment Corporation (DEC) PDP/11 mainframe system \citep{Childers1992}. The receiver architecture was reengineered to enable continuous coverage of the full 300 MHz bandwidth spanning the 1400–1700 MHz ``Water Hole” region of the radio spectrum. Spectral analysis was performed using 3000 frequency channels, each with a bandwidth of 100 kHz.

The data processing architecture shifted from static observations to an automated "frequency zoom" and tracking routine. The telescope’s physical feed horns were integrated onto a motorized tracking line called the ``SETI Eastwestern Railway". When the real-time software detected an anomalous narrowband signal rising through the sidereal transit, it immediately paused the automated survey. The processing system engaged tracking motors to move the feed horns along railroad tracks, locking onto the celestial position for about an hour to gather high-resolution spectral and temporal follow-up data before the source drifted out of range.

\subsection{Phase IV - 1995–1997 (SERENDIP)}

In its final operational phase, the Big Ear telescope’s analog front-end was integrated with a state-of-the-art digital signal-processing back-end engine \citep{Brown1997}. The Search for Extraterrestrial Radio Emissions from Nearby Developed Intelligent Populations (SERENDIP) system provided real-time analysis of four million 0.6 Hz channels spanning a 2.5 MHz bandwidth, with an integration time of 1.7 seconds. The SERENDIP III system, funded by NASA and developed at the University of California, Berkeley, was used to complete a 3,000-hour survey of the 1421.75–1424.25 MHz frequency band. An antenna pattern-matching algorithm was developed and implemented to reduce false alarms. This system functioned in parallel with LOBES throughout the final days of the telescope, until its decommissioning.

\subsection{Phase V - 2003–2025 (Argus)}

\begin{figure}
    \centering
    \includegraphics[width=1\linewidth]{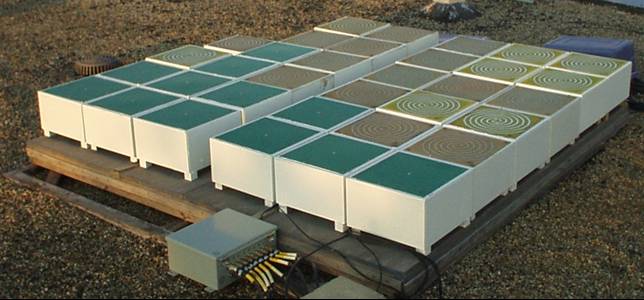}
    \caption{Argus was an experimental, all-sky radio telescope designed to provide an instantaneous field of view of the majority of the celestial sphere at L-band frequencies. The project was funded by the SETI Institute and private donations.}
    \label{fig:argus}
\end{figure}

Following the physical destruction of the Big Ear, the Ohio SETI Program transitioned from a single giant meridian telescope to an experimental all-sky software-defined radio architecture known as Project Argus \citep{1997abos.conf..623D,1998AAS...193.1117B,ellingson_argus_2008, ohioargus}. Rather than using analog mirrors to focus radio waves, Project Argus utilized a distributed planar array of 24 spiral elements of small, inexpensive, omnidirectional spiral antennas (\Cref{fig:argus}). Argus was operated from the campus of The Ohio State University.

The processing paradigm switched entirely to digital beamforming. Each antenna element digitized the raw incoming radio frequency voltage directly at the field point. These synchronized digital data streams were routed to a central computer cluster. By applying complex phase-delay algorithms to the combined data vectors in real time, the software acted as a ``Radio Camera," computationally synthesizing independent virtual antenna beams simultaneously. This architecture processes the entire visible hemisphere at once, ensuring that transient or short-duration celestial pulses are captured regardless of when or where they appear in the sky.

\section{Results}

\subsection{The Wow! Signal}

\begin{figure}
    \centering
    \includegraphics[width=1\linewidth]{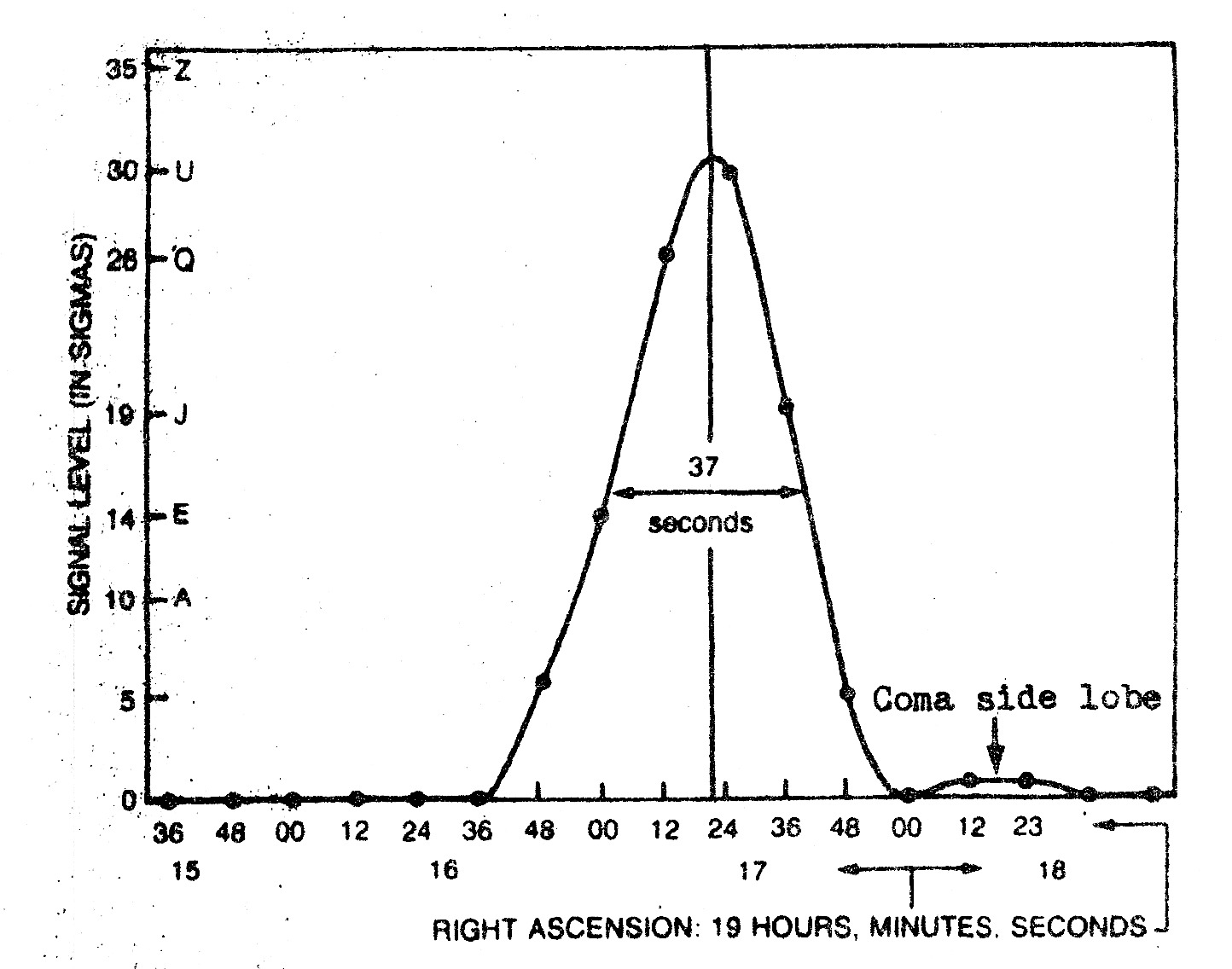}
    \caption{The Wow! Signal was detected on 15 August 1977, and this image constitutes its earliest known graphical representation, produced by John D. Kraus between September and October 1977. Source: The Big Ear Archives, Arecibo Wow!.}
    \label{fig:wow}
\end{figure}

The most prominent and famous detection of the Ohio SETI Program is the event that has become known as the Wow! Signal, observed in 1977 during Phase II \citep{kraus1979}. The name was unintentionally coined when astronomer Jerry R. Ehman wrote the exclamation "Wow!" in the margin of the computer printout upon noticing the striking data sequence. This signal exhibited all the expected characteristics of an extraterrestrial beacon. It was intense, spectral in nature, and matched the telescope's unique antenna pattern, which is consistent with a celestial point source farther away than the lunar distance (\Cref{fig:wow}). A review of public satellite data confirmed that no known satellites or deep-space missions were in the vicinity, and the signal's frequency was near the 1420 MHz hydrogen line, a band protected by international agreement from terrestrial and space-based communications.

One of the most perplexing features of the Wow! Signal was its appearance in only one of the telescope's two feed beams. The radio telescope was designed to receive two beams from the sky simultaneously (slightly offset in direction) and subtract them to cancel out terrestrial interference. A true celestial object should typically pass through both beams with a slight delay. However, the Wow! Signal was received only once, meaning it either turned off entirely after passing through the first beam, or turned on just after the first beam had passed. Following the detection, the observatory dedicated many days to observing the location of the signal, at \nobreakdash-27 degrees declination. Despite a hundred subsequent observations, the signal was never found again. The exact nature of the Wow! Signal remains a mystery.

\subsection{Other Signals}

Beyond the Wow! Signal, the 50-channel system, also during Phase II, recorded and archived over 40,000 signals of interest. The vast majority of these were isolated, narrowband pulses lasting between 10 to 30 seconds \citep{1985IAUS..112..305D}. Because they never repeated from exactly the same direction, they could not be attributed to a single source. However, spatial mapping of these transient signals revealed highly unusual, non-random astronomical distributions:

\begin{itemize}

    \item Galactic Latitude Dependency: The isolated pulses exhibited a distinct zone of avoidance along the galactic equator, showing a general anti-correlation between the number of detections and galactic latitude. In contrast, the signals showed areas of concentration above and below the galactic center, along the galactic north and south polar axes. Although this could theoretically be a complex instrumental artifact caused by the interaction of the galactic continuum radiation with the computer's automatic gain and baseline algorithms, resurveys of the area showed the phenomenon to be repeatable.

    \item Two Galactic Hot Spots: Statistical maps of the sky revealed two discrete areas near the galactic plane, exhibiting a much higher than average number of single-point detections.

    \begin{itemize}
    
    \item The Northern Area: Located slightly earlier than the galactic center in right ascension, this area was populated by a large number of low-level detections (92 isolated detections exceeding $5\sigma$, with a maximum of $19\sigma$). This area was observed in both September 1977 and May 1978, showing consistent behavior.

    \item The Southern Area: Located about two degrees south of the galactic center and later in right ascension, this hot spot was characterized by a smaller number of very high-intensity detections. The data revealed 41 isolated detections that exceeded $5\sigma$, 11 exceeded $20\sigma$, and the highest reached $30\sigma$. Surveyed in December 1977 and again in June 1978, the repetitive nature of this region across a six-month gap likely rules out localized terrestrial interference.
    
    \end{itemize}
    
\end{itemize}

\subsection{Serendipitous Astronomical Discoveries}

The design of the Ohio SETI Program was intentionally formulated to permit serendipitous detection of natural astrophysical phenomena in addition to the targeted search for SETI signals. A continuum 8 MHz receiver, tuned at the hydrogen line frequency, was also in operation in parallel during most observation surveys. This was the same receiver used in the original OSS of the 1970s. This receiver not only provided an independent diagnostic of telescope performance and contextual metadata for the SETI observations, but also facilitated systematic monitoring of radio sources and transient phenomena across a temporal range encompassing approximately three decades of the Big Ear operation. There are numerous transient events in these observations that require detailed investigation.

The telescope's surveys also yielded findings such as small cold hydrogen clouds. For example, in 1988, volunteer observer Tom Van Horne examined archival SETI records and identified a narrowband radio source that had been detected in early 1980. Unlike a point source, the emission appeared to be spatially extended. Its morphology and observing frequency were consistent with emission from neutral hydrogen, suggesting the presence of a diffuse hydrogen cloud. This object became informally known as the Van Horne Hydrogen Cloud in the observatory and was re-observed multiple times in 1993 as part of the LOBES survey. The Van Horne hydrogen cloud was subsequently identified as cores of the IV Arch, later designated IV21 and characterized in detail by Kuntz and Danly in 1996 \citep{1996ApJ...457..703K, vanhorne}. These clouds had already been observed 16 years earlier by the Big Ear, but unfortunately they were never reported.

\section{Conclusion}

The Ohio SETI Program fundamentally anticipated the modern rise of time-domain radio astronomy. By executing an uninterrupted, decades-long baseline of identical all-sky observations, the Big Ear created a highly unique temporal archive. This deep historical data set allows researchers to study long-term variability in the radio sky by comparing real-time measurements directly against a baseline of 30 years of 70\% of the sky, a cross-generational comparison impossible at any other observatory. This methodology is highly valuable not only for identifying potential artificial beacons but also for systematically hunting for variable radio sources and tracking slow, structural transformations in natural galactic or extragalactic features.

The signals captured by the Big Ear remain a captivating, unresolved chapter in astrophysics. Nearly fifty years after its detection, the Wow! Signal stands as the most compelling candidate for an extraterrestrial transmission ever recorded. Because it vanished mid-observation and never repeated despite dedicated follow-up sweeps, its definitive origin remains formally classified as unknown. The thousands of short-duration, narrowband pulses recorded near the galactic center and polar axes remain similarly ambiguous. Although they could point toward an unrecognized, localized astrophysical phenomenon, they cannot be conclusively verified due to their transient, non-repeating nature. They represent some of the earliest documented examples of radio transients, serving as an enigmatic precursor to modern discoveries such as Fast Radio Bursts (FRBs).

The Ohio SETI program operated predominantly through volunteer efforts and private donations, supplemented by limited grant funding from agencies such as NASA. In 1976, it became one of the earliest, if not the first, observational SETI initiatives to receive formal support from NASA. In the 1990s, a massive initiative was undertaken to digitize and transfer more than two million punch cards and magnetic tapes first onto floppy disks, and then onto modern digital media, in order to safeguard an irreplaceable record of our historical radio sky. This preservation project drew profound support from the Planetary Society and prominent scientists such as Carl Sagan.

The ultimate legacy of the Ohio SETI Program extends far beyond its raw data. As the longest-running continuous SETI search in history, it proved that breakthrough science could be sustained through sheer collaborative willpower and citizen-science devotion. Furthermore, the program pioneered crucial structural foundations for modern astrophysics. The transition from the massive, static physical reflectors of the Big Ear to the software-defined, all-sky architecture of Project Argus helped lay the conceptual groundwork for modern omnidirectional phased arrays.

\section{The Future}

Although the Big Ear radio telescope was decommissioned and physically dismantled several decades ago, its extensive archival data, long assumed to have been permanently lost, has in fact been preserved. In 2024, astronomers at the Planetary Habitability Laboratory at the University of Puerto Rico at Arecibo launched the ``Arecibo Wow!" project, an ambitious initiative aimed at revisiting the Ohio SETI Program's legacy \citep{arecibowow}. A primary objective of this project is to digitize, catalog, and safeguard all remaining data and documentation from the Big Ear telescope, preserving these historical records for future analysis. By utilizing modern digital signal-processing techniques, a massive leap forward from the manual, paper-based analysis of the 1970s, the project seeks to comprehensively re-examine the Wow! Signal and search for similar transient events in historical data from both the Big Ear and the Arecibo Observatory. So far, the Arecibo Wow! team has made significant progress in understanding the Wow! Signal, detailed in two publications:

\begin{itemize}

    \item Arecibo Wow! I: An Astrophysical Explanation for the Wow! Signal (2024): The team proposed a novel natural explanation, suggesting that the original Wow! Signal could have been the first recorded instance of an astronomical maser-like flare. In this scenario, an intense, short-lived radiation source, such as a magnetar flare or a soft gamma repeater, triggered a sudden maser flare or superradiant emission of the hydrogen line, briefly enhancing the brightness of a cold hydrogen cloud in the galaxy \citep{2024arXiv240808513M}.
    
    \item Arecibo Wow! II: Revised Properties of the Wow! Signal from Archival Ohio SETI Data (2025): By applying modern methods to the historical data, the team revised the signal's parameters. They measured a much higher peak flux density, a new location seven arcminutes off in the right ascension, and a different frequency. This revised frequency implied that the source had a substantially higher radial velocity than originally assumed. The study also reinforced an astrophysical source for the signal rather than terrestrial radio interference \citep{2025arXiv250810657M}.

\end{itemize}

The Arecibo Wow! project is extending the scientific legacy of the Big Ear by applying modern data-analysis techniques to one of the oldest long-term SETI archives ever assembled. A comprehensive analysis of the archive will likely require many years, and previously unrecognized signals or phenomena may still remain hidden within the data. Today, most of the observations collected by the Ohio SETI Program remain unexplored.

\vspace{1em}
\noindent \textbf{Acknowledgments}
\vspace{1em}

The continued safeguarding of the Big Ear radio telescope’s scientific contributions and historical heritage was made possible not only by the work of its former staff, but also by NAAPO’s long-term, dedicated engagement of numerous volunteers over many decades. NAAPO continues to maintain an online presence and ongoing communication. We also recognize the substantial role Marc Abel, Scott Horn, Tom Hanson, and Robert H. Gray have played in preserving both the scientific work and the historical record of the Ohio SETI Program.

\bibliography{references}
\bibliographystyle{iaulike}

\end{document}